\documentstyle[12pt,epsf]{article}

\renewcommand{\bar}[1]{\overline{#1}}

\newcommand{\etal}{{\em et al.}}

\begin{document}
\begin{flushright}
SLAC-PUB-7553\\
DOE/ER/40561-330-INT97-19-07\\
DOE/ER/41014-16-N97 \\
July 1997
\end{flushright}
\bigskip\bigskip

\centerline{{\Large\bf Is $J/\psi$-Nucleon Scattering
            Dominated}}
\centerline{{\Large\bf by the Gluonic van der Waals Interaction?}
             \footnote{\baselineskip=14pt
     Work supported in part by the Department of Energy, contract
     DE--AC03--76SF00515.}}

\bigskip
\centerline{Stanley J. Brodsky}
\vspace{8pt}
  \centerline{Stanford Linear Accelerator Center}
  \centerline{Stanford University, Stanford, California 94309}
  \vspace{22pt}
\centerline{Gerald A. Miller
  \footnote{\baselineskip=14pt
  Permanent address: Department of Physics,
            Box 351560, University of Washington,\hfill\break
            \hbox{\hskip .25in}Seattle, Washington 98195-1560}}
\vspace{8pt}
   \centerline{Stanford Linear Accelerator Center}
   \centerline{Stanford University, Stanford, California 94309}
   \centerline{and}
   \centerline{National Institute for Nuclear Theory}
   \centerline{Box 35150, University of Washington}
   \centerline{Seattle, Washington 98195-1560}
 \vspace{22pt}

\begin{abstract}
The gluon-exchange contribution to $J/\psi$-nucleon scattering is shown
to yield  a sizeable scattering length of about -0.25 fm, which is
consistent with the sparse
available data. Hadronic corrections to gluon exchange which are generated by
$\rho\pi$ and $D\bar D$ intermediate states of the $J/\psi$
are shown to be negligible. We also propose a new method to
study $J/\psi$-nucleon elastic scattering in the reaction $\pi^+
d\to J/\psi\; p\; p$.
\end{abstract}
\vfill
\centerline{Submitted to Physics Letters B.}
\vfill
\newpage

One of the novel features that quantum chromodynamics brings to strong
interaction physics is the concept of a gluonic van der Waals potential, the
interaction arising from the exchange of two or more gluons between
color-singlet hadrons. The color van der Waals potential is expected to be the
dominant potential in the case of the scattering of hadrons without common
quarks, such as in the interaction of heavy quarkonium states with hadrons or
nuclei at low energies. As in quantum electrodynamics, the QCD van der Waals is
attractive, and in principle it could lead to molecular-like bound states
of charmonium with nuclei\cite{brod1,dave}.   Unlike QED,  the QCD van der
Waals potential has finite range, rather than the power-law fall-off
characteristic of the exchange of massless neutral gauge fields.\cite{Sucher}

It is clearly very interesting to study the theoretical foundations and
the empirical consequences of the van der Waals potential.  In an illuminating
paper, Luke, Manohar, Savage\cite{mls} have
shown that the essential features of the low energy interaction between
heavy quarkonium and nucleons or nuclei can be determined directly from the
operator product expansion.  In their analysis  the coupling of
multiple gluons to a small-size quarkonium bound state is given by the
quarkonium color electric polarizability.  The coupling of the gluons to the
large-size nucleon or nucleus depends on one term  proportional to
the momentum fraction carried  by gluons and a second term normalized to the
nucleon or nuclear mass. The dominant
low energy interaction at small relative velocity corresponds to scalar
exchange.
The gluon exchange potential can then lead to resonances or even bound states in
quarkonium-hadron or quarkonium-nuclear interactions.  Such novel states
could be
seen for example as kinematical peaks in the decay of the $B$ meson in the
$\bar
p \Lambda J/\psi$ final state.\cite{bn}

The main purpose of this letter is to demonstrate explicitly that the QCD
van der
Waals potential as characterized by its scattering length is indeed much more
important than the meson-exchange forces in
$J/\psi$-nucleon interactions which arise from the coupling of
charmonium to hadronic intermediate states. We also point out that the QCD van
der Waals interaction can be conveniently studied experimentally in the
highly-constrained reaction
$\pi d
\to J/\psi  p p$.

Indirect information on the interactions of the $c \bar
c$ system with nucleons can also be obtained from studies of charm
production at threshold or, via unitarity, the behavior of $p p$ elastic
scattering in the charm threshold region. Indeed the strong increase of
the polarization asymmetry $A_{NN}$ observed in Ref.~\cite{spin} in large
CM angle proton-proton scattering at $\sqrt s \simeq 5$ GeV has been
attributed to the strong interactions between charm anti-charm
configurations arising  in the intermediate state interacting with nucleons
at low relative  velocity\cite{brod2}. It would clearly be very useful
to verify these  physical features
from direct measurements of the $J/\psi$-nucleon interaction.

We begin by deriving the scattering length for the QCD van der Waals
potential, starting with the
Luke, Manohar, Savage LMS two-gluon exchange
calculation of the forward invariant amplitude in
first Born approximation:
\begin{equation}
{\cal M}_{\rm fwd}=4M_\psi\;M^2 {c_E\over\Lambda_Q^3}\left[
{3\over 4} \; V_2(\Lambda_Q)+
{2\pi \over \beta_Q \alpha_s(\Lambda_Q)}\right].
\label{fscat}
\end{equation}
Here $M$ is the nucleon mass and  $V_2$ is  the gluon momentum
fraction in the nucleon. (We take $V_2=0.5$ \cite{field} at the low
momentum transfer scale relevant here.) LMS assumed the heavy quark limit,
in which
the size of the heavy quarkonium
system, $r_B\sim 1/m_Q$ is much smaller  than the inverse of the QCD
scale $\Lambda_{QCD}^{-1}$, and in which
the $Q\bar Q$ system can be can be
approximated as a  Coulomb bound state. Peskin \cite{peskin}  found
\begin{equation}
{c_E\over\Lambda_Q^3}
= {14\pi\over 27}r_B^3
\label{aepeskin}
\end{equation}
where $r_B$ is the Bohr radius of  the $1s$ state. The LMS analysis is
a rigorous prediction of QCD, and  it is completely model independent in the
limit $m_Q\rightarrow\infty$. Although the validity of the Coulomb
approximation for the $J/\psi$ may not be completely reliable, this
result should provide a good estimate for $c_E$.  The LMS
computation shows unambiguously that the potential is attractive.

The parameters appearing in Eq.~(\ref{fscat}) must be specified
to obtain numerical results. The Bohr
radius of the $J/\psi$ has been  determined in the model-independent
analysis of Quigg and Rosner~\cite{qr} as $r_B^{-1}=750 $ MeV. The
value of the strong coupling constant at low momentum transfer scales
$Q\sim r_B^{-1}$ can be determined in the $\alpha_V$ scheme from
various exclusive processes \cite{brpr}. A convenient parameterization
which freezes the coupling at low scales is
\begin{equation}
{2\pi\over \beta_Q\alpha_V(Q)}= {1\over 2}
\ell n \left({Q^2+4\;m_g^2\over \Lambda_V^2}\right),
\end{equation}
with  $m_g^2=0.2$ GeV$^2$ and $\Lambda_V=0.16$ GeV. We now use the
relation $f_B={-{\cal M}_{fwd}\over 8\pi (M+M_\psi)}$ \cite{pdg} to
obtain the first Born forward scattering amplitude with the traditional
normalization used in potential theory. At threshold $f_B=a_B$, the
Born approximate scattering length. The numerical evaluation of
Eq.~(\ref{fscat}) using  the above parameters gives $a_B=-0.19$ fm.

We can go beyond the Born approximation to obtain the full scattering
length $a$ by assuming a specific form for the potential $V(r)$  in the
Schroindinger equation. The relation between the $a_B$ and $V(r)$ is:
\begin{equation}
a_B=2\mu \int_0^\infty\; dr\;r^2\;V(r),\label{ab}
\end{equation}
where $\mu$ is the reduced mass of the $J/\psi$-nucleon system. Since
the QCD van der Waals potential is of finite range, we shall assume a
Gaussian shape: $V(r)=-V_0 e^{-r^2/R^2}$ with $R$=0.8 fm chosen to
account for  the finite sizes of the nucleon, the $J/\psi$ and the range of
their interaction. 
Equation~(\ref{ab}) implies then that the potential depth is $V_0=23$ MeV,
which is quite large.  Solving the Schroindinger equation then gives
$a=-0.24$ fm, corresponding to a total cross section $\sigma_{\psi N}=
4\pi a^2$ of about   7 mb at threshold.   This is somewhat higher than
the values $\sigma_{\psi N}$=3.5$\pm$ 0.8 mb $\pm$ 0.5 mb determined
from nuclear  $J/\psi$ photoproduction at 20 GeV \cite{tot}. However the
low energy $J/\psi$-nucleon interaction corresponds to scalar exchange,
implying a cross section which decreases as the energy is increased
from threshold. Thus the low-energy value of 7 mb is not inconsistent with
the higher energy determination. This number does not depend much on the shape
of the potential, for a fixed value of $a_B$, as long as
the range is held fixed.

One possible signal for a strong $c \bar c$ interaction with nucleons
is the abrupt rise of the polarization asymmetry $A_{NN}$ observed in
large CM angle proton-proton elastic scattering at $\sqrt s \simeq 5 $
GeV near the charm threshold. De Teramond and collaborators\cite{guy} 
have argued that reproducing the 
strength of the polarization asymmetry $\sqrt{s}$ in the vicinity of the
threshold for $c\bar c$ production (about  10 GeV) 
requires an effective
scattering length 
of about the same size as our result.
We note that the estimate of Ref.~\cite{guy} includes
the interactions of  an ensemble of quarkonium states whose
interactions with nucleons could 
be different than that of the $J/\psi$.

\vspace{.5cm}
\begin{figure}[htbp]
\begin{center}
\leavevmode
\epsfbox{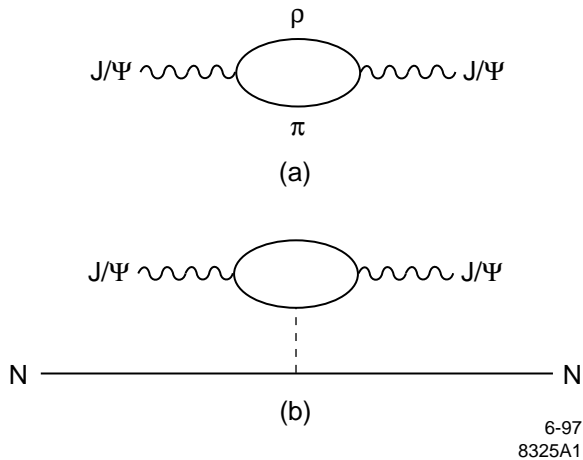}
\end{center}
\caption[*]{Influence of $\rho\pi$ intermediate states on $M_\psi$.
(a)  free $J/\psi$ (b) typical graph for
$J/\psi$ in the presence of a nucleon}
\label{fig1}
\end{figure}

Does multiple-gluon exchange really dominate over all other strong
interaction effects in $J/\psi$-nucleon scattering?  It is natural
to consider the effects involving the exchange of pions between the $J/\psi$
and the nucleon. 
Isospin conservation prevents the exchange of a single pion, but
the two-pion exchanges generated by contact interactions which take a
$J/\psi$ into a $J/\psi \pi\pi$ state are allowed. In the LMS
formalism, such terms owe their existence to non-vanishing masses of the
light quarks. Their size is of order 
$({m_\pi\over 4\pi f_\pi})^2\approx  1\%$
of the terms in Eq.~(\ref{fscat})\cite{mls}.

But there could be other types of hadronic interactions.
To investigate the possibilities we look at the
hadronic width  of the $J/\psi$.   The $J/\psi$ decay to $\rho\pi$ has a
remarkably large 1.28 \% branching fraction of the 87 KeV
width \cite{pdg}, an effect which has been attributed to the intrinsic
charm Fock states of the $\rho$ and $\pi$ mesons \cite{bk}. Thus the
$J/\psi$ could interact with a nucleon via virtual $\rho\pi$
interactions,  as in Fig.~1b. In order to estimate the strength of such
meson-exchange contributions,  we postulate a hadronic Lagrangian of
the form
\begin{equation}
{\cal L}_{\rho\pi}= g \bar\psi_\mu \mbox{\boldmath$\rho^\mu$}\cdot
\mbox{\boldmath$\pi$},
\end{equation}
the simplest interaction term consistent with isospin and space-time symmetries.
The coupling $g$ is readily determined from the width to the $\rho\pi$
channel:
\begin{equation}
\Gamma_{\rho\pi}= {g^2\over 8\pi}
(3 + {p_1^2\over M_\rho^2}){p_1\over M_\psi^2},
\end{equation}
where $p_1$ is the relative momentum of the $\rho\pi$ system, $p_1=0.47
M_\psi$. Then $g=\epsilon M_\psi$ with $\epsilon=1.7\times 10^{-3}$.

Given a value of $g$ we may compute the scattering amplitude due to the
graphs of Fig.~1. We use relativistic time-ordered perturbation theory
to obtain the $J/\psi$-nucleon interaction where the time ordering
corresponds to the intermediate state $\rho\pi$ plus the nucleon.
Our  procedure will be to first consider the free-space amplitude of
Fig.~1a which contributes to the mass of the $J/\psi$ and then see how
this mass shift is modified by the presence of the nucleon. Let
$V_\psi^{(0)}$ be the shift in the mass of the $J/\psi$ due to the
$\rho\pi$ channel:
\begin{eqnarray}
V_\psi^{(0)}&=&\langle\psi|V_{\rho\pi}{1\over E-H_0+i\epsilon}V_{\rho\pi}
|\psi\rangle\nonumber\\
&=&
{\Gamma_{\rho\pi} \over 2\pi} {M_\psi\over p_1}
Re \int_0^\infty {q^2 dq\over E_\pi E_\rho}
{1\over (E-E_\pi-E_\rho+i\epsilon)} \label{s1}
\end{eqnarray}
where $E=M_\psi $ for the on-shell $J/\psi$. Here $H_0$ accounts for
the sum of the relativistic kinetic energies
$E_\pi=\sqrt{q^2+M_\pi^2}$, $E_\rho=\sqrt{q^2+M_\rho^2}$ of the $\rho$
and $\pi$, and $V_{\rho\pi}$ is the interaction Hamiltonian derived
from ${\cal L}_{\rho\pi}$. Note that $Im
V_\psi^{(0)}=-{\Gamma_{\rho\pi}\over 2}$. The mass shift is modified in
the presence of a nucleon:
\begin{equation}
V_\psi=\langle\psi,N|V_{\rho\pi}
{1\over E-H_0-V_N+i\epsilon}V_{\rho\pi}|\psi,N\rangle
\label{s2}\end{equation}
where $V_N$ is the strong interaction potential arising between the
$\rho$ and the $\pi$ and the nucleon.

The $J/\psi$-nucleon interaction $V(\psi,N)$ that we seek is given by
the difference $ V(\psi,N)=Re \left(V_\psi-V_\psi^{(0)}\right)$ as in
the Furry picture. The evaluation of Eqs.~(\ref{s1}) and (\ref{s2}) is
complicated; however, for our purposes we can make  an estimate by
taking $V_N$ to be a constant. Thus to first order in $V_N$ we find
\begin{equation}
V(\psi,N)
\approx {V_N \over M_\psi-M_\rho-M_\pi}
{\Gamma_{\rho\pi} \over 2\pi}{M_\psi\over p_1}.
 \end{equation}
The small value of $\Gamma_{\rho\pi}$ (= 1.1 KeV) causes $V(\psi,N)$
to be  of the order of $10^{-5}$ MeV, which is entirely negligible.

\vspace{.5cm}
\begin{figure}[htbp]
\begin{center}
\leavevmode
\epsfbox{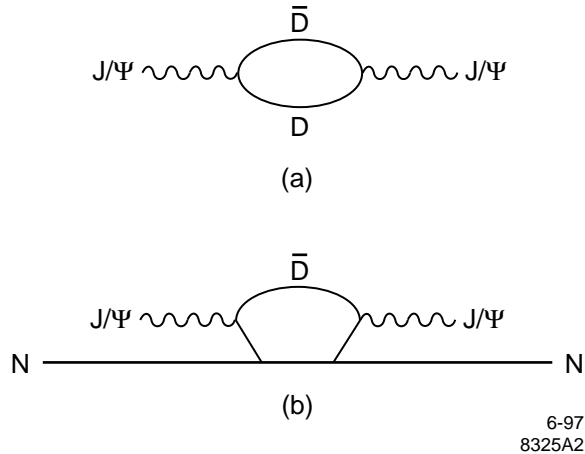}
\end{center}
\caption[*]{Influence of $\bar D D$ intermediate states on $M_\psi$.
(a) free $J/\psi$ (b)$J/\psi$ in the presence of a nucleon, sample  graph.
The intermediate baryon carries charm.}
\label{fig2}
\end{figure}

Another type of term is shown in Fig.~2, which is driven by the
coupling of the $J/\psi$ to a pair of $D$ mesons. The $J/\psi$ is
stable with respect to this decay, but it could lead to an interaction
with nucleons through the virtual intermediate states.  A significant
interaction could be possible even at low energies since the anti-light
quark could be absorbed and the charmed quark added to the nucleon to
make an intermediate $\Lambda_c$. We again use a hadronic effective
Lagrangian
\begin{equation}
{\cal L}_{D\bar D}= i\;g_c \psi^\mu\left(\bar{{\bf{D}}} \partial_\mu\cdot
{\bf{D}}-{\bf{D}}\partial_\mu\cdot\bar{{\bf{D}}} \right).
\end{equation}
We cannot obtain $g_c$ from the width of the $J/\psi$, so
we shall extrapolate from  the decay of the $\phi$ using an interaction
\begin{equation}
{\cal L}_{K\bar K}= i\;g_s \phi^\mu\left(\bar{\bf{K}} \partial_\mu\cdot
\bf{K}-\bf{K}\partial_\mu\cdot\bar{\bf{K}} \right).
\end{equation}
which also involves a vector meson decaying into two pseudoscalar
particles. We take $g_s$ from the decays $\phi\to K^0\bar K^0$ and
$\phi\to K^+K^-.$  A simple evaluation gives ${g_s^2\over 4\pi}=1.71$.

Obtaining the relation between $g_c$ and $g_s$ is
the next task. The key feature is the small-sized nature of the $J/\psi$.
We recall  that the small size of the region involved in $c\bar c$ annihilation
was the crucial ingredient in the Appelquist-Politzer\cite{ap}
explanation of the
very small hadronic width of the $J/\psi$. This is because
the
amplitudes for gluons emitted by a charmed and nearby
anti-charmed quark tend to
cancel.
Taking the emission of the light-quark pair to be governed by two gluon-
exchange, as in the LMS example, leads to a transition amplitude given by
the matrix element of the   operator $r$ representing the distance
between the heavy quarks. The power arises from the small
size of the initial system. A related example, \cite{mc,fs}
is the ratio of the decay amplitudes for
$\Upsilon'$ and $\psi'$ to decay to their ground states
and two pions which is governed by the ratio of the
mean square radii\cite{mc}. In that case, both the initial ($\Upsilon'$,
$\psi'$)  and  final ($\Upsilon$,
$\psi$) states involve small systems, so that one obtains two powers of $r$.
We also note that
the ratio of the same  decays of the $\psi'$ and the $\rho'$ is
very small\cite{pdg}.
Another  related example occurs
when considering the coupling constant for pions to interact with the
point-like configuration of the nucleon\cite{fs}. In that case,
the coupling  also varies as the area  of the point-like configuration.
Thus  we expect that
\begin{equation}
g_c\approx g_s {R_\psi\over
R_\phi}
\end{equation}
Using the root-mean-square radii $R_\psi=0.11$ fm \cite{qr} and
$R_\phi=0.4$  fm\cite{ph} leads to
\begin{equation}
{g_c^2\over 4\pi}\approx 0.13. \label{gc}
\end{equation}

We shall evaluate the contribution to the $J/\psi$-nucleon interaction
coming from Fig.~2 by examining how  the shift in $M_\psi^2$ of
Fig.~2a, $\Delta M_\psi^2(M_D^2),$ changes in the presence of a nucleon,
modelling the effect of the nucleon with the replacement $M_D\to M_D+V(D,N).$
The $D$-nucleon interaction $V(D,N)$ is treated as a constant.
The resulting contribution to the $\psi$-N potential $V(\psi,N$
is then given by
\begin{equation}
V(\psi,N)={\partial \Delta M_\psi^2
\over \partial M_D^2}{M_D\over M_\psi}V(D,N), \label{difff2}
\end{equation}
with
\begin{equation}
\Delta M_\psi^2(M_D^2)=
i\; g_c^2\; \int {d^4k\over (2\pi)^4}\; {(\epsilon(\lambda)\cdot (2k-P))^2
\over (k^2-M_D^2 +i \epsilon)\;((k-P)^2-M_D^2 +i \epsilon) },
\end{equation}
where $P$ is the four momentum of the $J/\psi$, and $\epsilon (\lambda)$ its
polarization vector. We
 average over the $J/\psi$ polarization $\lambda$ and combine the energy
denominators to obtain:
\begin{equation}
\Delta M_\psi^2(M_D^2)=
i\; g_c^2\;\int_0^1\;dx \int {d^4k\over (2\pi)^4}\; {k^2
\over (k^2-\Delta(x))^2 }, \label{denom}
\end{equation}
where $\Delta(x)\equiv M_D^2-M_\psi^2\; x(1-x)+i \epsilon>0.$
This term can be  renormalized via the dimensional regularization procedure,
corresponding to the removal of an infinite term independent of $M_D^2$
which does not contribute to $V(\psi,N)$ of Eq.~(\ref{difff2}). In particular,
we  replace the factor of $k^2$ appearing in the
numerator of Eq.~(\ref{denom}) by $\Delta(x)$.
Then
\begin{equation}
{\partial \Delta M_\psi^2\over \partial M_D^2}=
2\;i\; g_c^2\; \int_0^1\;dx\int {d^4k\over (2\pi)^4}\; {\Delta(x)
\over (k^2-\Delta(x))^3 }={1\over 4\pi} {g_c^2\over 4\pi},
\end{equation}
and
\begin{equation}
V(\psi,N)={1\over 4\pi} {g_c^2\over 4\pi}
 {M_D\over M_\psi}V(D,N)= 0.006\; V(D,N).\label{diff2}
\end{equation}
We may obtain
an upper limit for $V(D,N)$ by assuming that the $D$-nucleon and
$\pi$-nucleon interactions  are similar. The $\pi$-N
interaction corresponds \cite{ew} to a strength of approximately 1 MeV
near threshold and about 20 MeV for pions of 100 MeV. Thus,
the estimated value of $V(\psi,N)$ is less than about 0.1 MeV and
negligible compared to the 23 MeV from multiple gluon exchange.

The net conclusion is that the QCD van der Waals potential for
$J/\psi$-nucleon interactions from gluon exchange is strongly dominant
over meson-exchange interactions.

\vspace{.5cm}
\begin{figure}[htbp]
\begin{center}
\leavevmode
\epsfbox{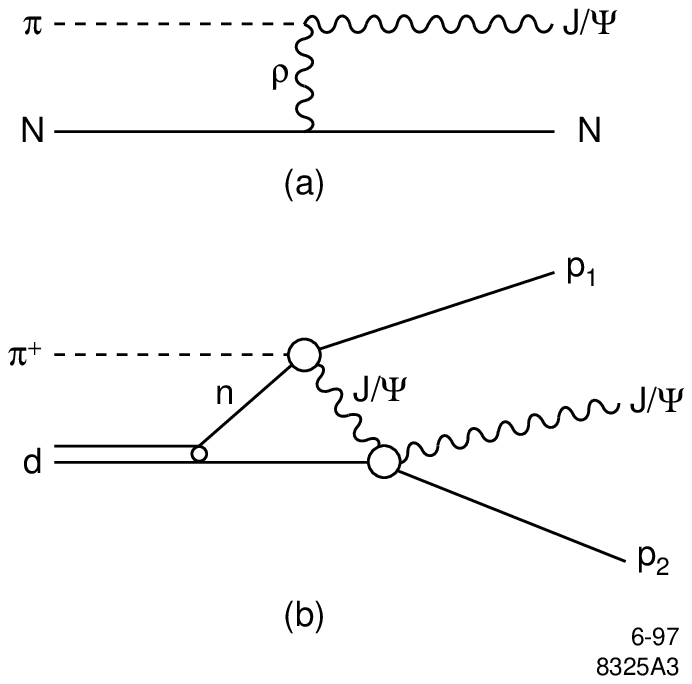}
\end{center}
\caption[*]{(a) The production process $\pi^+n\to J/\psi p$.
(b) The production process $\pi^+d\to J/\psi p_1\; p_2$.}
\label{fig3}
\end{figure}

It is clearly interesting and important to measure $J/\psi N$
scattering. The traditional and most straightforward method is to analyze the
nuclear dependence of $J/\psi$ photoproduction.\cite{first} However, the
$\rho\pi$ interaction offers the opportunity to measure the $J/\psi N$
interaction in a totally exclusive situation.  Consider the process $\pi^-
p\to\psi n$ which proceeds by $\rho$ exchange as shown in Fig. ~3. The cross
section for this process is easily evaluated. In the limit $s\gg\;M_\psi^2$
\begin{equation}
{d\sigma_{\pi N\to\psi N}\over dt}={1\over 2} {g_{\rho N}^2\over 4\pi}
\epsilon^2
{F^2(t)\over (t-M_\rho^2)^2},
\end{equation}
where the $\rho$-nucleon coupling constant ${g_{\rho N}^2\over 4\pi}=1$,
and $F(t)$ is the $\rho$-nucleon form factor. We find ${d\sigma_{\pi
N\to\psi N}\over dt}(t=0)=$ 1.6 nb  GeV$^{-2}$. This may be compared
with the photoproduction cross section of 10 nb GeV$^{-2}$ observed in
$\gamma p\to \psi p$ \cite{first}, and thus it is not too small to be observed.
The total cross section for this $J/\psi$ production process is
approximately 1 nb. In principle, one can also use this reaction to search for
$J/\psi-N$ resonances or even $J/\psi-N$  bound states.

More generally one can study multiple-scattering corrections to the process
$\pi  N \to \psi N$  in a nuclear target and thus  measure how the
$J/\psi$ scatters on a nucleon. For example, consider the deuteron target
process
$\pi^+ d\to J/\psi+p_1+p_2$ as shown in Fig.~3b. The process $\pi^+ $ +
neutron $\to  J/\psi$ plus proton of momentum $p_1$ is followed by a
$J/\psi $ + proton scattering. If the momentum $p_2$ of the second
proton is greater than about 300 MeV/c,  such an event can only be
produced by a scattering \cite{fgmss}, because
the deuteron wave function does
not have appreciable support for such high momenta. The signature of
$J/\psi$ production is a monoenergetic peak in the missing mass
obtained by measuring the two final state nucleons, one of momentum
close to that of the beam and the other of greater than 300 MeV/c. Such
fixed target measurements would enable the first  measurement of
$J/\psi$-nucleon elastic scattering at near threshold energies.

The results presented here demonstrate that the exchange of two gluons
in the scalar channel dominates elastic $J/\psi-N$ scattering. The most
logical hadronic mechanisms which might be competitive are interactions
involving $\rho\pi$ or $D\bar D$ intermediate states. The contributions to the
quarkonium-nucleon  potential  are found to be very small compared to the
predicted color van der Waals strength.

We find that the cross section for
exclusive
$J/\psi$ production in pion-nucleon collisions
$\pi^+ d\to J/\psi+p_1+p_2$ is not negligible, enabling a
very clean measurement of the magnitude and momentum dependence of
$J/\psi$-N elastic scattering at low energies. Such measurements could
test the first-principle theoretical predictions for the  QCD van der Waals
interaction.

{\bf Acknowledgments}

This work is partially supported by the USDOE. G. A. Miller  thanks
the SLAC theory group, the National Institute for Theoretical Physics
and the Adelaide center for their hospitality. We also thank M. Savage for a
useful discussion.

\end{document}